\documentclass[10pt,
               aps,
               pre,
               twocolumn,
               showpacs, 
               groupedaddress]{revtex4-2}
\usepackage[utf8]{inputenc}
\usepackage{graphicx}
\usepackage{amsmath}

\usepackage[dvipsnames,table]{xcolor}
\usepackage[caption=false]{subfig}
\usepackage{amsmath, amsfonts, amssymb}
\usepackage{hyperref}
\usepackage{tikz}
\graphicspath{{images/}}

\makeatletter
\newcommand*{\rom}[1]{\expandafter\romannumeral #1}
\makeatother

\allowdisplaybreaks

\begin{document}

\title{Chiral active gyrator: Memory induced direction reversal of rotational motion}
\author{S Deion}
\affiliation{Department of Physics, University of Kerala, Kariavattom, Thiruvananthapuram-$695581$, India}

\author{F Adersh}
\affiliation{Department of Physics, University of Kerala, Kariavattom, Thiruvananthapuram-$695581$, India}

\author{M Sahoo}
\email{jolly.iopb@gmail.com}
\affiliation{Department of Physics, University of Kerala, Kariavattom, Thiruvananthapuram-$695581$, India}

\date{\today}

\begin{abstract}
We theoretically explore the dynamics of a chiral active Ornstein–Uhlenbeck particle confined in a two-dimensional anisotropic harmonic trap. The particle is driven by chirality and is coupled to two orthogonal heat baths, potentially at two different temperatures. Using both analytical approach and numerical simulation, we explore the rotational dynamics of the particle in both viscous and viscoelastic environments. While the particle is suspended in a viscoelastic bath, characterized by a finite memory time scale, we interestingly observe that even in the absence of a temperature gradient, the angular momentum changes its sign as a function of the memory timescale, reflecting the direction reversal of rotational motion of the particle, and it is solely due to the interplay between memory and chirality. 
This direction reversal is a distinct memory-induced phenomenon and does not occur in the viscous limit. Moreover, increasing (or decreasing) the temperature gradient shifts the magnitude of the angular momentum further into the negative (or positive) direction, selecting a unique direction of rotation for the particle throughout the activity-memory parameter space.
However, in the viscous limit, the direction reversal of the rotational motion is still possible but occurs only across a neutral line in parameter space, along which the contributions from both chirality and thermal anisotropy to the net angular momentum exactly cancel. These results highlight a distinct mechanism for directional control in active systems, with memory enabling reversal phenomena unique to viscoelastic media.
\end{abstract}

\maketitle

\section{INTRODUCTION}
Active matter is a rapidly expanding field that spans physics, chemistry, biology, and materials science~\cite{Ramaswamy2017active,gompper2020, needleman2017, magistris2015intro}. 
Unlike passive systems, where motion typically arises from external driving forces or environmental asymmetries, active systems are composed of self-driven units that continuously consume energy to generate autonomous motion. As a result, they exhibit intrinsically out-of-equilibrium dynamics~\cite{marchetti2013hydrodynamics}. 
This self-propelling nature allows them to move, organize, and interact collectively without the need for external control~\cite{bechinger2016active}, often resulting in emergent structures and complex dynamical phases that have no analog in equilibrium systems~\cite{henkes2020dense, zottl2016, mallory2018active}. 
Examples range from biological entities such as bacteria~\cite{Peruani2012bacteria}, sperm cells~\cite{Schoeller2020CollectiveSperm}, and flocking birds~\cite{Mora2016LocalEquilibriumflocking} to artificial systems such as synthetic self-propelled colloids~\cite{zottl2016} and Janus particles~\cite{walther2013janus}.
A hallmark feature of such systems is the spontaneous breaking of time-reversal symmetry, often resulting in persistent motion and exotic transport phenomena~\cite{martin2021statistical, dabelow2019irreversibility}.

Among active particles, chiral active particles (CAPs) are particularly intriguing because of the inherent rotational bias in their self-propulsion~\cite{Liebchen2022chiral}. The chirality introduces a persistent angular component in the particle motion, which arises from intrinsic particle asymmetry. Consequently, these particles can both self-propel and self-rotate~\cite{lowen2016chirality, sahala2023self-propulsion_nodate}.
The motion of CAPs can be described by introducing chirality into the well-established active particle frameworks. For instance, chirality can be incorporated in frameworks such as Active Brownian Particles (ABPs)~\cite{ten2011ABP, Malakar2020ABP2d, caprini2021collective} or Active Ornstein–Uhlenbeck Particles (AOUPs)~\cite{bonilla2019active,martin2021statistical,caprini2021inertial,caprini2019active,sevilla2019generalized}, via torque-like effects or by coupling the translational and rotational degrees of freedom~\cite{caprini2018confinement, sahala2023self-propulsion_nodate}.
The chiral activity or chiral self-propulsion gives rise to a rich spectrum of phenomena across a broad scale~\cite{caprini2018confinement, dominik2024nonreciprocal_chiral, mijalkov2013sorting}.
At the single-particle level, chiral swimmers follow circular or helical trajectories determined by the strength of chirality~\cite{sahala2023self-propulsion_nodate}.
At the collective level, they display vortex and spiral flows~\cite{Caprini2024vortex}, flocking and structure formation~\cite{PhysRevLett.119.058002}, as well as self-organization and phase separation~\cite{liebchen2017collective}.
However, most of these studies consider purely viscous or memoryless environments.

In reality, many natural and synthetic media do not respond instantaneously to deformation. Instead, the internal stress, i.e., the restoring force per unit area generated by the deformation of the medium, relaxes over a finite timescale. This means that the medium is viscoelastic and retains a memory of how it was deformed in the past, giving rise to a history-dependent (non-Markovian) dynamics \cite{Goychuk2009subdiffusion,muhsin2021orbital}. 
Viscoelasticity is common in complex fluids such as polymer solutions, gels, mucus, and the cytoplasm of living cells~\cite{Bhat12viscoelastic,GUIGAS2007biovisco}. 
In such media, the interplay between viscous dissipation and elastic storage of energy due to deformation of the medium gives rise to long-lived temporal correlations in the particle dynamics~\cite{Goychuk2009subdiffusion, grimm2011brownian}. These correlations can be described by incorporating a friction kernel in the Langevin framework~\cite{das2023enhanced,sevilla2019generalized}. Remarkably, even achiral particles can trace circular trajectories purely due to memory effects in viscoelastic media~\cite{narinder2018memory_induced}. 
In general, non-Markovian environments can either amplify or oppose activity, strongly influencing the trajectory and directionality~\cite{solano2016dynamics, schamel2014nano, LI2021104655, shen2011swimmervisco, ferrer2021fluid,Gomez2015colloidalvisco,das2023enhanced}. Recently, it has been observed that viscoelastic memory influences the translational dynamics of chiral active particles in viscoelastic environments~\cite{Kobayashi2024_chiral_viscoelastic}. These developments highlight the growing importance of understanding how chirality and viscoelastic memory interplay in more complex settings.

Beyond viscoelastic effects, environmental asymmetries, including anisotropic confinement, temperature gradients, and external forces, further modulate the nonequilibrium features of active transport.
Even a passive Brownian particle, when coupled to two heat baths and confined in an anisotropic harmonic potential, can exhibit gyration around the potential minimum~\cite{mancois2021two_temp,dotsenko2013two,filliger2007brownian, nascimento2020memory}.   
Interesting effects emerge for an active particle in such systems when subjected to synthetic Lorentz forces~\cite{kumar2012classical,muhsin2021orbital,abdoli2022tunable}. It results in trapped diamagnetism~\cite{muhsin2024magneto}, and allows engine- or heat pump-like operation~\cite{adersh2025activemagnetogyratortunable}.
Moreover, the interplay between chirality and external asymmetries such as potential anisotropies and temperature gradients can give rise to bulk accumulation~\cite{caprini2018confinement, caprini2023chiral_ou} and other interesting transport behaviors like ordered oscillations~\cite{lei2023collective} and rectified transport~\cite{Bao2019chiral_twotemp}.
Furthermore, it would be interesting to explore the gyrating behavior of CAPs in viscoelastic environments.
The complex interplay of chirality, viscoelastic memory, confinement, and temperature gradient is expected to enrich the dynamics.

Motivated by this, we investigate the dynamics of a CAP confined in an anisotropic harmonic well, immersed in a viscoelastic medium, and subjected to a temperature gradient. We derive analytical expressions for the mean angular momentum and observe that, unlike in the purely viscous case, viscoelastic memory can induce the direction reversal of rotational motion of the particle even in the absence of temperature gradient. We note that anisotropy of the confining potential is necessary for such direction reversal to occur. Furthermore, by tuning the temperature gradient away from zero, we demonstrate that the directionality of angular momentum can be controlled and stabilized into a single preferred direction of rotation. We map out the conditions for this memory-induced direction reversal of rotational motion and contrast them against the viscous limit, thus highlighting the unique feature of non-Markovian mechanism at play.

\section{MODEL AND METHOD}\label{sec:model}
We consider an inertial chiral active particle characterized by its chirality strength $\Omega$ moving in the $x$–$y$ plane. The system is confined in an anisotropic harmonic potential $V$, which has the form
\begin{equation}
V(x, y) = \frac{1}{2}k(x^2 + y^2) + \alpha x y.
\label{eq:potential}
\end{equation}
Here, the parameter $k$ represents the stiffness of the potential, while $\alpha$, which has the same dimensionality as that of $k$, determines the shape of the potential by quantifying the potential anisotropy. The potential $V$ becomes isotropic for $\alpha = 0$.
The particle is driven by a chiral self-propulsion modeled as an Ornstein-Uhlenbeck process, which drives the system out of equilibrium and breaks time-reversal symmetry.
In addition, the medium is viscoelastic with finite memory effects and has two thermal heat baths at different temperatures. 
The position vector ${\bf r} = x {\bf \hat{i}} + y {\bf \hat{j}}$ evolves according to the generalized Langevin dynamics ~\cite{igor2012visco, sevilla2019generalized}, 
given by
\begin{equation}
m \ddot{\mathbf{r}}(t) = -\int\limits_{-\infty}^{t} f(t-t') \dot{\mathbf{r}}(t') \, dt' - \boldsymbol{\nabla} \boldsymbol{V}+ \boldsymbol{\xi}(t) + \boldsymbol{\eta}(t).
\label{eq:model}
\end{equation}
The first term on the LHS of Eq.~\eqref{eq:model} represents inertial force, while the first term on the RHS captures the viscoelastic drag via a friction kernel $f(t - t')$, defined as~\cite{das2023enhanced, sevilla2019generalized, fa2008study, raikher2013brownian}
\begin{equation}
f\left(t-t'\right)=
\frac{\gamma}{2} \delta(t - t') + \frac{\gamma}{2t_c'}e^{-\frac{t - t'}{t_c'}}.
\label{eq:kernel}
\end{equation}
Here, $\gamma$ is the viscous or friction coefficient of the medium. $t_c'$ denotes the viscoelastic memory time scale, and it is the duration of time up to which there exists finite memory in the medium. The thermal noise $\boldsymbol{\eta}(t) = \eta_x(t) \hat{i} + \eta_y(t) \hat{j}$ satisfies zero mean $\langle \eta_i(t) \rangle = 0$ and the generalized fluctuation–dissipation relation $\langle \eta_i(t) \eta_j(t') \rangle = \delta_{ij} T_i f(t - t')$, with $i, j \in \{x, y\}$. Here $T_x$ and $T_y$ are along the $x$ and $y$ directions, respectively.

The stochastic term $\boldsymbol{\xi}(t) = \xi_x(t) \hat{i} + \xi_y(t) \hat{j}$ is the active force that captures the chiral self-propulsion, and hence follows a chiral extension of the Ornstein-Uhlenbeck process~\cite{martin2021statistical, nguyen2022active, sahala2023self-propulsion_nodate} given by
\begin{equation}
\dot{\boldsymbol{\xi}}(t) 
= -\frac{1}{t_c}\boldsymbol{\xi}(t) 
+ \xi_0 \sqrt{\tfrac{2}{t_c}} \, \boldsymbol{\zeta}(t) 
+ \Omega \, \boldsymbol{\xi}(t) \times \hat{\bf z}.
\label{eq:chiral_noise}
\end{equation}
Where $\xi_0$ is the amplitude of the active force, $t_c$ is the activity timescale, and $\Omega$ is the chirality strength. The term $\boldsymbol{\zeta}(t)$ in Eq.~\eqref{eq:chiral_noise} is a Gaussian white noise with zero mean and unit variance, i.e., $\langle \zeta_i(t) \rangle = 0$ and $\langle \zeta_i(t) \zeta_j(t') \rangle = \delta_{ij} \delta(t - t')$.
While modeling the active force $\boldsymbol{\xi}(t)$, chirality is introduced through the rotational term $\Omega \boldsymbol{\xi}(t) \times \hat{\mathbf{z}}$~\cite{caprini2018confinement,caprini2023chiral_ou}, which causes the active force vector to rotate in the $x$–$y$ plane around the $z$-axis. This term acts perpendicular to $\boldsymbol{\xi}(t)$ and changes its orientation without affecting its magnitude, thereby inducing circular or spiral trajectories. The unit vector $\hat{\mathbf{z}}$ points along the positive $z$-axis, and the parameter $\Omega$ quantifies the rate of this active rotation and serves as a measure of the chirality strength of the particle. 
The active force $\boldsymbol{\xi}(t)$ satisfies the properties 
\begin{equation}
\langle \xi_i(t) \rangle = 0,\nonumber
\end{equation}
{\centering
and
\begin{equation}
\langle \xi_i(t)\, \xi_j(t') \rangle = \xi_0^2 \delta_{ij} 
\exp\!\left( -\frac{|t - t'|}{t_c} \right) 
\cos\!\big[\Omega (t - t')\big],
\label{eq:xi_corr}
\end{equation}}
which reflects exponentially decaying temporal chiral activity modulated by a chiral oscillation at frequency $\Omega$.
In the limiting cases $t_c \to 0$ or $\Omega \to \infty$, the active force $\boldsymbol{\xi}(t)$ vanishes, i.e., $\boldsymbol{\xi}(t) \to 0$, and the system reduces to that of a passive Brownian particle.
Throughout this work, we use dimensionless variables and set the Boltzmann constant $k_B = 1$. We introduce the parameters $\Gamma = \frac{\gamma}{m}$, $\omega_0 = \sqrt{\frac{k}{m}}$ and define the temperature gradient $\Delta T$ as $T_x - T_y$, with $\Delta T = 0$ corresponding to same temperature of both heat baths. 

\begin{figure*}
    \centering
    \includegraphics[width=0.85\linewidth]{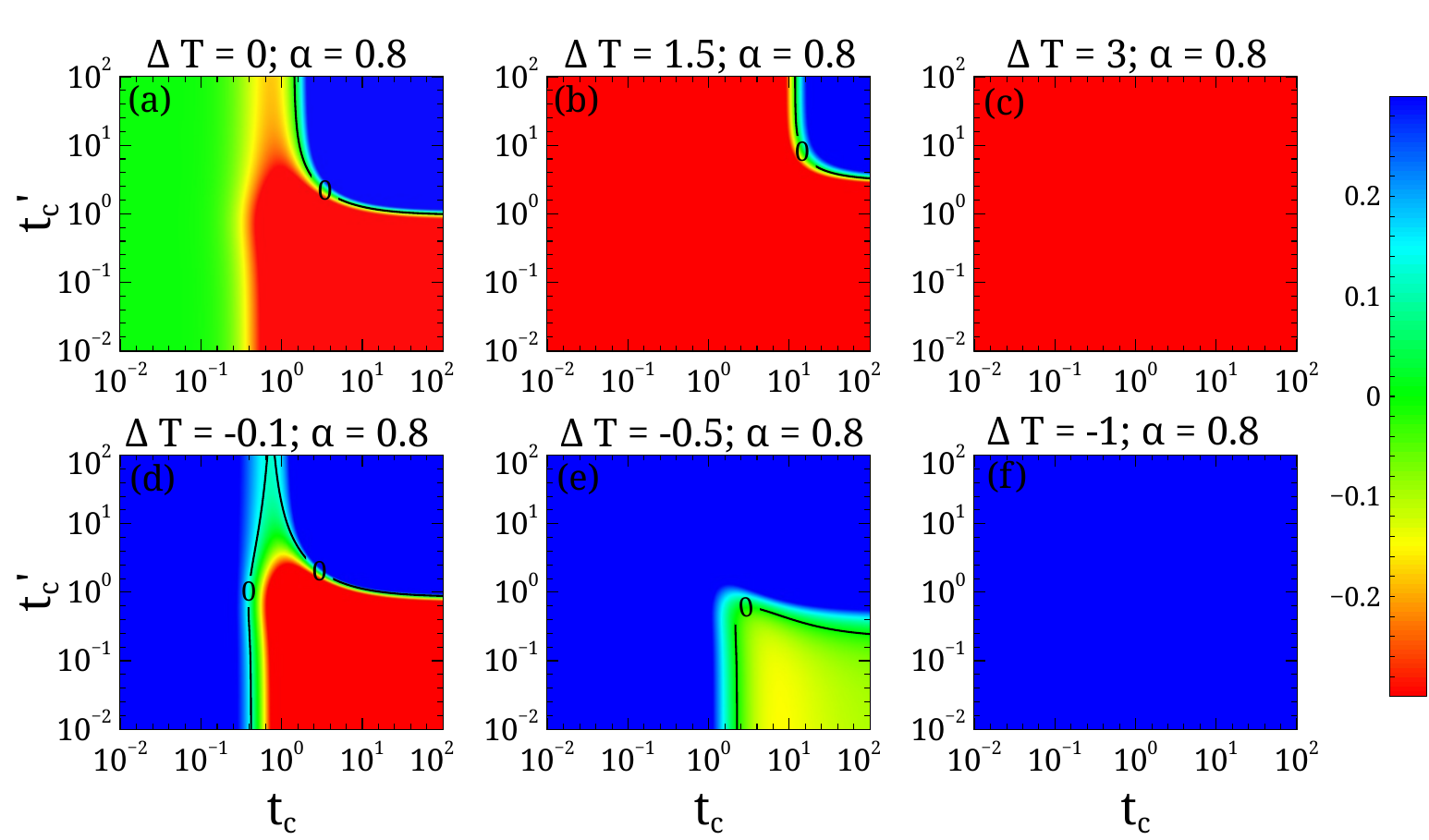}
    \caption{The 2D parametric plot of $\langle \omega \rangle$ [Eq.~\eqref{eq:omega_gen}] as a function of $t_c$ and $t_c'$ is shown in (a) for $\Delta T = 0$, (b) for $\Delta T = 1.5$, (c) for $\Delta T = 3$, (d) for $\Delta T = -0.1$, (e) for $\Delta T = -0.5$, and (f) for $\Delta T = -1$. The color map represents the value of $\langle \omega \rangle$. The common parameters are $\alpha = 0.8$, $m = \Omega = \xi_0 = \omega_0 = \Gamma = 1$.
}
    \label{fig:omega_vs_tc_tcp}
\end{figure*}
 
The intrinsic chirality of the particle imparts an effective rotational component to its self-propulsion~\cite{sahala2023self-propulsion_nodate}. Simultaneously, the presence of both temperature gradient and potential anisotropy induces gyration of the particle across the potential minimum~\cite{mancois2021two_temp}. Hence, the combined effect of both chiral self-propulsion and gyration drives the system out of equilibrium and produces an effective resultant behavior that manifests itself as a net rotational bias to the particle motion. This rotational bias can be quantified by measuring the effective angular momentum. The steady-state average of the corresponding angular momentum can be defined as
\begin{equation}
    \langle \omega \rangle = \lim_{t \to \infty}|\langle {\bf r} \times {\bf v} \rangle|,
    \label{eq:M_def}
\end{equation}
where ${\bf v} [= \dot{\bf r}]$ is the instantaneous velocity of the particle.

We have simulated the dynamics governed by Eq.~\eqref{eq:model} along with the evolution of the active force described by Eq.~\eqref{eq:chiral_noise}, using the Euler--Maruyama integration scheme~\cite{kloeden1992numerical}.
The simulations were carried out with a time step of $10^{-2}$, and a total of $10^{3}$ timesteps were taken to ensure that the system reaches a steady state. Subsequently, ensemble averages were computed over $10^{5}$ independent realizations to obtain statistically reliable results.

\section{RESULTS AND DISCUSSION}\label{sec:result}

\subsection{Chiral Active Viscoelastic Gyrator}
When the particle is suspended in a viscoelastic environment, the dynamics is given by Eq.~\eqref{eq:model}. With the help of the steady state correlation matrix method as discussed in Appendix~\ref{sec:app_A}, we have calculated $\langle {\bf r} \times {\bf v} \rangle$ [Eq.~\eqref{eq:omega_relation}]. Then the exact solution of $\langle \omega \rangle$ [Eq.~\eqref{eq:M_def}] of the particle is given by
\begin{widetext}
\begin{equation}
\begin{split}
\langle \omega \rangle = &
- \frac{2 \Delta T \alpha \Gamma}{D} \left[
2 {t'_c}^4 \alpha^2 
- m^2 \left( 2 + t'_c \left( \Gamma + t'_c \omega_0^2 \right) \right) 
       \left( 2 + t'_c \left( \Gamma + 2 t'_c \omega_0^2 \right) \right)
\right] \\
& - \frac{2 t_c^2 \xi_0^2 \Omega}{m D} \left[
\frac{N}{
S + (m \omega_{0}^2 - \alpha)\left(
S' + 4 t_c^4 (m \omega_{0}^2- \alpha) E
\right)} +
\frac{M}{
S + (m \omega_{0}^2 + \alpha)\left(
S' + 4 t_c^4 (m \omega_{0}^2 + \alpha) E
\right)}
\right],
\end{split}
\label{eq:omega_gen}
\end{equation}
\end{widetext}
where $S$, $S'$, $D$, $E$, $M$, and $N$ are given in Appendix~\ref{sec:app_B}. The two terms of $\langle \omega \rangle$ [Eq.~\eqref{eq:omega_gen}] are controlled by two distinct physical mechanisms. The first term is non-zero finite only when both $\alpha$ and $\Delta T$ are non-zero. This reflects the gyration of the particle around the potential minimum that emerges from the combined effect of potential anisotropy and temperature gradient~\cite{mancois2021two_temp}. This term vanishes if either $\alpha$ or $\Delta T$ becomes zero. In contrast, the second term arises due to the chiral self-propulsion and is non-zero finite for non-zero values of both $\Omega$ and $t_{c}$. This term vanishes if either $t_{c}$ or $\Omega$ becomes zero. Moreover, the persistence of viscoelastic memory (finite values of $t_c'$) affects both the terms, leading to a resultant angular momentum. Even though $t_c'$ influences both terms, its effect is not uniform. The first term shows only a minimal change with an increase in $t_c'$, whereas the second term exhibits a more pronounced dependence, resulting in noticeable modulation of the net angular momentum and hence the overall rotational behavior. Depending on the relative strengths of both persistent chiral activity and viscoelastic memory, the system exhibits qualitatively different dynamical behavior across regimes of the activity-memory parameter regimes. In particular, we observe that the persistence of viscoelastic memory can amplify or suppress the resultant rotational motion of the particle through the second term, leading to a sign reversal in $\langle \omega \rangle$ that separates regions of opposite rotational directions of the particle.

In Fig.~\ref{fig:omega_vs_tc_tcp}, we have shown the two-dimensional (2d) plot of $\langle \omega \rangle$ as a function of both $t_c$ and $t_c'$ for a fixed $\alpha$ and for different values of $\Delta T$. When $\Delta T = 0$ [Fig.~\ref{fig:omega_vs_tc_tcp}(a)], the first term in Eq.~\eqref{eq:omega_gen}, which is proportional to $-\alpha\Delta T$, vanishes. As a result, $\langle \omega \rangle$ is purely due to the second term, which is due to the combined effect of both viscoelastic memory and chiral self-propulsion.
This provides a clear setting for isolating the combined effect of chiral activity and memory-induced dynamics. In this case, $\langle \omega \rangle$ remains negative for most of the $t_c$-$t_c'$ parameter regime. In the limit $t_c \to 0$, the second term tends to zero, leading to a vanishing $\langle \omega \rangle$. However, for finite $t_c$, $\langle \omega \rangle$ is negative for lower values of $t_{c}'$. As $t_c'$ increases, the term $N$ containing $t_c'$ undergoes a sign reversal, leading to a transition from negative to positive $\langle \omega \rangle$. This shows a clear memory-induced sign change for $\langle \omega \rangle$, implying a memory-induced direction reversal of the rotational motion. The system thus exhibits co-existence of opposite angular momentum, separated by a neutral line along which $\langle \omega \rangle = 0$. The locus of all points on this neutral line satisfies the equation
\begin{equation}
\begin{split}
&\Delta T \alpha \gamma \Big[
2 {t'_c}^4 \alpha^2 
- m^2 ( 2 + t'_c (\Gamma + t'_c \omega_0^2)) \\
&\hspace{14.5em}\times ( 2 + t'_c (\Gamma + 2 t'_c \omega_0^2))
\Big] \\
&\quad= - \frac{2 t_c^2 \xi_0^2 \Omega}{m}
\Biggl[\frac{N}{%
S + (m \omega_{0}^2 - \alpha)(
S' + 4 t_c^4 (m \omega_{0}^2- \alpha) E
)}\\
&\qquad\qquad\qquad
+ \frac{M}{%
S + (m \omega_{0}^2 + \alpha)\bigl(
S' + 4 t_c^4 (m \omega_{0}^2 + \alpha) E
\bigr)}\Biggr].
\end{split}
\label{eq:omega_zero}
\end{equation}
With an additional increase in the value of $\Delta T$ from zero to positive values [Fig.~\ref{fig:omega_vs_tc_tcp}(b)], the first term in Eq.~\eqref{eq:omega_gen} becomes negative and the total $\langle \omega \rangle$ increases in the negative direction. This leads to a shrinkage of the positive $\langle \omega \rangle$ region observed. For even higher $\Delta T$ values [Fig.~\ref{fig:omega_vs_tc_tcp}(c)], the first term dominates completely, resulting in a net negative angular momentum ($\langle \omega \rangle < 0$) across the entire parameter regimes. On the other hand, with a decrease in $\Delta T$ from zero to negative values, the first term becomes positive and starts to dominate in the lower $t_{c}$ regimes. The second term becomes relevant only for large $t_c$ values. Like before, due to the contribution from viscoelastic memory effects, this second term undergoes a sign change depending on the values of $t_{c}'$. This results in the appearance of two coexisting regions of positive and negative $\langle \omega \rangle$, separated by two neutral lines [Fig.~\ref{fig:omega_vs_tc_tcp}(d)].
With further decrease in $\Delta T$ values as in Fig.~\ref{fig:omega_vs_tc_tcp}(e), the positive contribution from the first term increases. This results in the shrinkage of the negative $\langle \omega \rangle$ region observed.
For sufficiently large negative $\Delta T$ [Fig.~\ref{fig:omega_vs_tc_tcp}(f)], the first term dominates the entire expression of $\langle \omega \rangle$, leading to a completely positive $\langle \omega \rangle$ throughout the entire $t_{c}-t_{c}'$ parameter regimes.
This indicates that by tuning the value of $\Delta T$ away from zero towards more positive (or negative) values, the system can be driven into a completely negative (or positive) $\langle \omega \rangle$ regime, eliminating the coexistence and selecting a unique direction for the rotational motion of the particle.
Similar behavior can be observed when $\alpha$ is taken as negative and $\Delta T$ is increased from a negative value towards a positive value through zero.
This is because changing the signs of both $\alpha$ and $\Delta T$ has no effect on the expression of $\langle \omega \rangle$ in Eq.~\eqref{eq:omega_gen}. That is, $\langle \omega \rangle(-\alpha, -\Delta T) = \langle \omega \rangle(\alpha, \Delta T)$. 
Moreover, with further decrease in the value of $\Gamma$, the coexisting region is found to persist even for larger values of $\Delta T$, before it eventually disappears. 
In the following sections, we discuss the nature of net angular momentum in the limiting cases where the resultant rotational motion is dominated either by the gyration or the chiral self-propulsion. This further helps us in understanding the general complex interplay between these two physical processes.

\begin{center}
{\textit{Case 1: Pure chiral self-propulsion ($\Delta T=0$)}}
\end{center}
Here, we consider the case where there is no gyration of the particle. This corresponds to a situation when either $\Delta T = 0$ or $\alpha = 0$, i.e., one anisotropy is absent. In this setting, we consider $\Delta T=0$, and the expression for $\langle \omega \rangle$ [Eq.~\eqref{eq:omega_gen}] becomes \\
\scalebox{0.95}{
\parbox{\linewidth}{
\begin{equation}
\begin{aligned}
\langle \omega \rangle =\ 
- \frac{2 t_c^2 \xi_0^2 \Omega}{m D} 
\left[
\frac{N}{
S + (m \omega_{0}^2 - \alpha)\left(
S' + 4 t_c^4 (m \omega_{0}^2- \alpha) E
\right)} \right. \\[6pt]
\left. + \frac{M}{
S + (m \omega_{0}^2 + \alpha)\left(
S' + 4 t_c^4 (m \omega_{0}^2 + \alpha) E
\right)}
\right].
\end{aligned}
\label{eq:omega_gen_deltaT0}
\end{equation}
}
}

\begin{figure}
    \centering
    \includegraphics[width=1\linewidth]{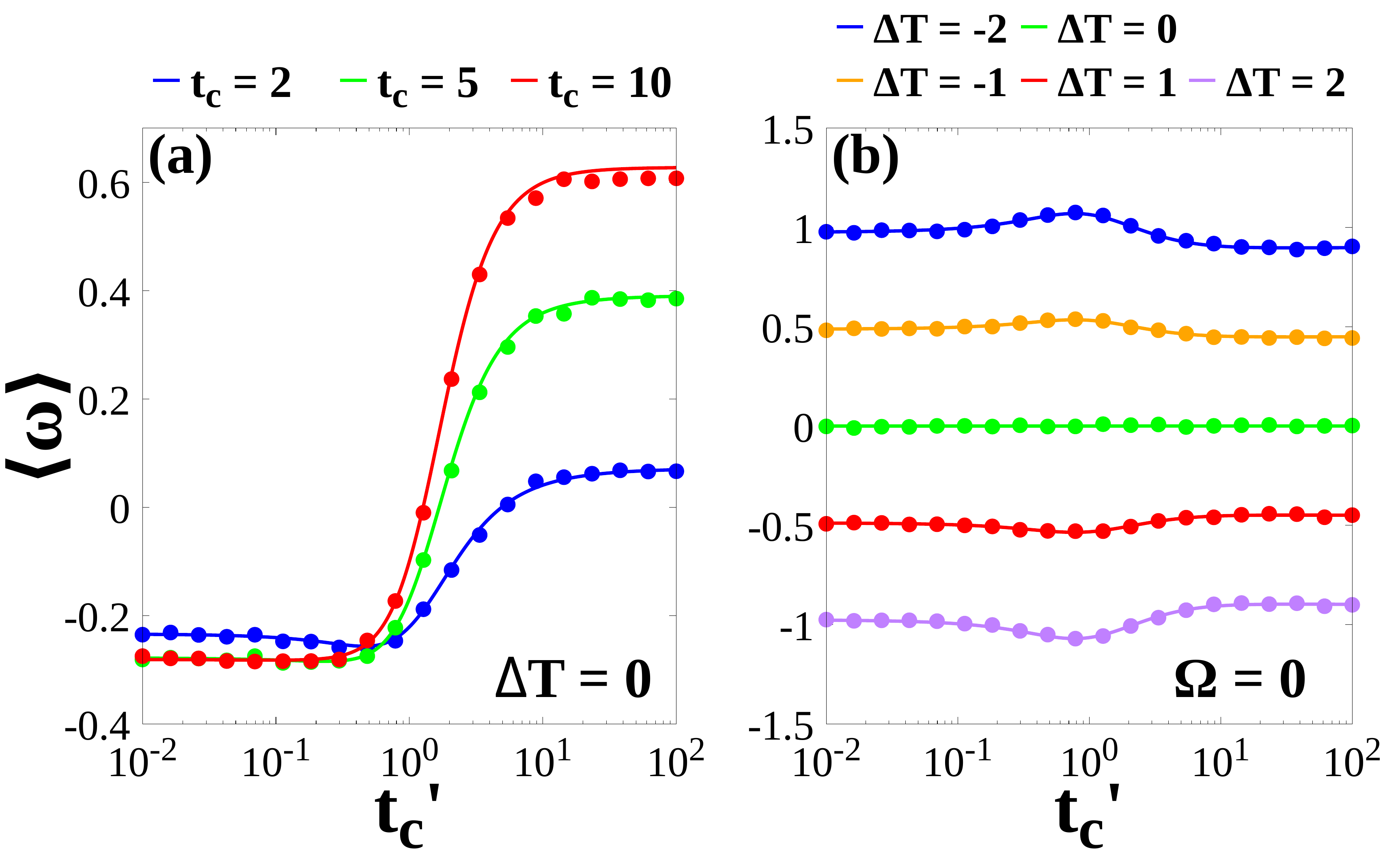}
    \caption{$\langle \omega \rangle$ [Eq.~\eqref{eq:omega_gen_deltaT0}] as a function of $t_c'$ is shown in (a) for $\Delta T = 0$, $\Omega = 1$ and different values of the $t_c$, and in (b) for $\Omega = 0$ and various values of $\Delta T$. The parameters used are $m = \Gamma = \omega_0 = \xi_0 = 1$ and $\alpha = 0.8$. The solid line represents the analytical results, and the dots represent the results obtained
    from numerical simulation.
    }
    \label{fig:omega_vs_tcprime}
\end{figure}

From Eq.~\eqref{eq:omega_gen_deltaT0}, it is confirmed that in the absence of a temperature gradient, $\langle \omega \rangle$ depends intricately on $t_c$, $\Omega$, and $t_c'$. The angular dynamics stem from the interplay between chiral activity and viscoelastic memory. In Fig.~\ref{fig:omega_vs_tcprime}(a), using both analytical results and numerical simulation, we explore the behavior of $\langle \omega \rangle$ as a function of $t_c'$ for a fixed $\Omega$ and for different values of $t_c$.
In the $t_{c}' \rightarrow 0$ limit, i.e., in the persistence of very short viscoelastic memory, $\langle \omega \rangle$ [Eq~\eqref{eq:omega_gen_deltaT0}] reduces to
\begin{widetext}
\scalebox{0.97}{
\parbox{\linewidth}{
\begin{equation}
\begin{split}
\langle \omega \rangle =
-\frac{t_c^2 \xi_0^2 \Omega}{\alpha^2 + m^2 \Gamma^2 \omega_0^2} \Bigg[
    \frac{
      m (\alpha + t_c^2 \alpha \Omega^2)
      + m t_c^2 \Gamma^2 (\alpha + m \omega_0^2)
      - t_c (t_c \alpha - 2 m \Gamma)(\alpha + m \omega_0^2)
    }{
      m^2 t_c^4 \Omega^4 
      + \left(m + t_c^2 \alpha + m t_c (\Gamma + t_c \omega_0^2)\right)^2
      + m t_c^2 \Omega^2 \left(-2 t_c^2 \alpha 
        + m \left(2 + t_c (\Gamma (2 + t_c \Gamma) - 2 t_c \omega_0^2)\right)\right)
    } \\
  \qquad\qquad\qquad - 
    \frac{
      m (\alpha + t_c^2 \alpha \Omega^2)
      + m t_c^2 \Gamma^2 (\alpha - m \omega_0^2)
      + t_c (t_c \alpha + 2 m \Gamma)(\alpha - m \omega_0^2)
    }{
      m^2 t_c^4 \Omega^4 
      + \left(m - t_c^2 \alpha + m t_c (\Gamma + t_c \omega_0^2)\right)^2
      + m t_c^2 \Omega^2 \left(2 t_c^2 \alpha 
        + m \left(2 + t_c (\Gamma (2 + t_c \Gamma) - 2 t_c \omega_0^2)\right)\right)
    }
\Bigg].
\end{split}
\label{eq:omega_gen_deltaT0_tcp0}
\end{equation}
}
}
\end{widetext}

From the above equation, it is clear that in $t_{c}' \rightarrow 0$ limit, $\langle \omega \rangle$ is negative and becomes independent of $t_{c}'$. This corresponds to the left plateau seen in Fig.~\ref{fig:omega_vs_tcprime}(a).
A small increase in $t_c$ value results in $\langle \omega \rangle$ becoming more negative, reflecting faster rotation. However, further increase of $t_c$ doesn't add any significant impact on $\langle \omega \rangle$ in the lower $t_{c}'$ regime. For a fixed $t_{c}$ value, $\langle \omega \rangle$ shows a negative constant value in the lower $t_{c}'$ regime, increases monotonically with $t_{c}'$ in the intermediate $t_{c}'$ regimes, and then saturates to a constant value for very large $t_{c}'$ values. The intermediate increase of $\langle \omega \rangle$ with $t_{c}'$ is more pronounced with increase in value of $t_{c}$. 
Thus, in the intermediate regime of $t_{c}'$, the increase of $t_{c}$ results in a gradual crossover of $\langle \omega \rangle$ from negative to positive values, indicating a reversal in the effective direction of rotational motion.
Moreover, for very long persistence of viscoelastic memory in the medium, i.e., in $t_c' \rightarrow \infty$ limit, $\langle \omega \rangle$ approaches again to a constant value, and this constant value increases with increase in value of $t_{c}$ and saturates for very large $t_{c}$ values.

\begin{center}
    {\textit{Case 2: Pure gyration ($\Omega=0$)}}
\end{center}

 Next, we consider the case of pure gyration of the particle in a viscoelastic medium, where all angular motion originates purely due to the combined effect of thermal and potential anisotropies. In this configuration, i.e., in the absence of chiral motion ($\Omega=0$), the $\langle \omega \rangle$ [Eq.~\eqref{eq:omega_gen}] simplifies to
\begin{equation}
\begin{aligned}
\langle \omega \rangle
&= - \frac{2 \Delta T \alpha \Gamma}{D} \big[
    2 {t'_c}^4 \alpha^2 
    - m^2 \left( 2 + t'_c \left( \Gamma + t'_c \omega_0^2 \right) \right)
    \\
&\hspace{9.5em}
       \times \left( 2 + t'_c \left( \Gamma + 2 t'_c \omega_0^2 \right) \right)  \big].
\end{aligned}
\label{eq:omega_gen_omega0}
\end{equation}
From Eq.~\eqref{eq:omega_gen_omega0}, it is to be noted that $\langle \omega \rangle$ does not depend on $t_c$, which confirms that the activity timescale is irrelevant in this limit.
In Fig.~\ref{fig:omega_vs_tcprime}(b), we analyze the behavior of $\langle \omega \rangle$ [Eq.~\eqref{eq:omega_gen_omega0}] as a function of $t_c'$ for different values of $\Delta T$.
It is observed that for $\Delta T = 0$, $\langle \omega \rangle$ becomes zero for the entire regime of $t_c'$, indicating that the particle does not undergo rotational motion. 
For non-zero finite values of $\Delta T$, $\langle \omega \rangle$ becomes finite and exhibits a weakly non-monotonic dependence on $t_c'$ in the intermediate $t_{c}'$ regime. In the $t_c' \to 0$ limit, Eq.~\eqref{eq:omega_gen_omega0} reduces to
$\langle \omega \rangle = \frac{-\Delta T \alpha \Gamma}{ \alpha^2 + m^2 \Gamma^2 \omega_0^2}$ and in the $t_c' \rightarrow \infty$ limit, $\langle \omega \rangle$ reduces to $\langle \omega \rangle = \frac{-2 \Delta T \alpha \Gamma}{ 4 \alpha^2 + m^2 \Gamma^2 \omega_0^2}$. Hence, in both limits, $\langle \omega \rangle$ is independent of $t_{c}'$. 
For the intermediate regime of $t_{c}'$, $\langle \omega \rangle$ shows a minimum for a positive $\Delta T$ and the minimum value decreases with increasing $\Delta T$ values. Similarly, $\langle \omega \rangle$ shows a maximum for a negative $\Delta T$ and increases with a decrease in $\Delta T$. 
By changing the sign of the product $\alpha \Delta T$, the rotational motion of the particle can be tuned in the opposite direction throughout the whole $t_{c}'$ regime.
However, unlike the case of pure chiral motion, pure gyration does not exhibit a sign reversal of $\langle \omega \rangle$ with $t_c'$. This highlights that viscoelastic memory alone cannot induce rotational reversal in the absence of chirality but can slightly enhance or suppress $\langle \omega \rangle$ depending on its coupling to other anisotropies in the system.

\begin{figure}
    \centering
    \includegraphics[width=1.015\linewidth]{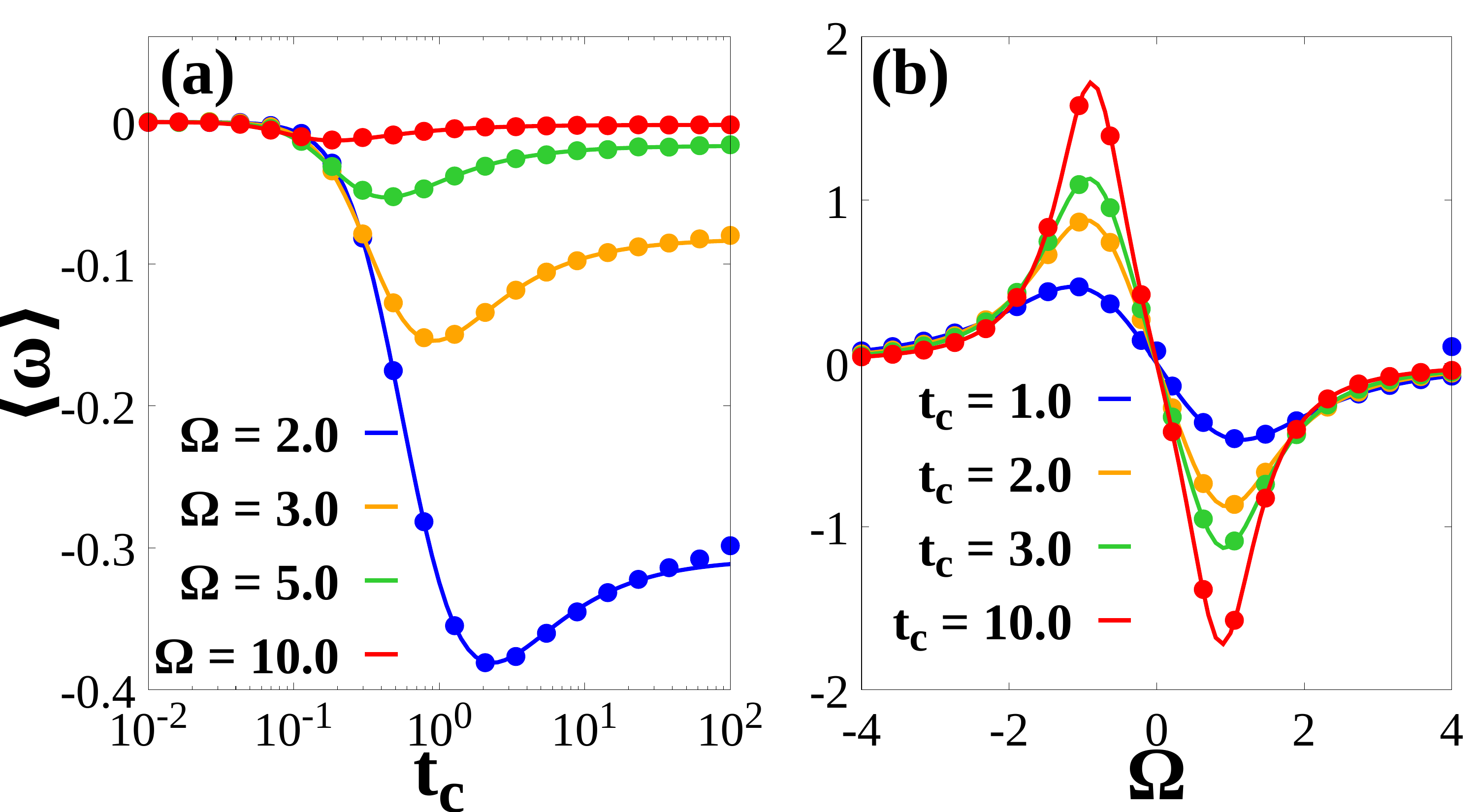}
    \caption{$\langle \omega \rangle$ [Eq.~\eqref{eq:omega_viscous_a0}] as a function of the $t_c$ for different values of $\Omega$ is shown in (a), and as a function of $\Omega$ for various fixed values of $t_c$ is shown in (b). The other common parameters used are $m = \Gamma = \xi_0 = \omega_0 = 1$. The solid line represents the analytical results, and the dots represent the results obtained from numerical simulation.}
    \label{fig:omega_viscous_a0} 
\end{figure}

\subsection{Chiral Active Viscous Gyrator}

In the limit $t_c' \to 0$, the friction kernel in Eq.~\eqref{eq:kernel} becomes $f(t - t') = \gamma\delta(t-t')$. With this, the system reduces to the case of a chiral active viscous gyrator and in this limit, the dynamics of the particle [Eq.~\eqref{eq:model}] can be written as
\begin{equation}
m \ddot{\boldsymbol{r}}(t) = -\gamma \dot{\boldsymbol{r}}(t) - \boldsymbol{\nabla V } + \boldsymbol{\xi}(t) + \boldsymbol{\eta}(t).
\label{eq:viscous_chiral_mass}
\end{equation}
The expression for $\langle \omega \rangle$ [Eq.~\eqref{eq:omega_gen}] in the limit $t_c' \to 0$, reduces to
\begin{widetext}
\scalebox{0.89}{
\parbox{\linewidth}{
\begin{equation}
\begin{split}
\langle \omega \rangle =
\frac{-\Delta T\, \alpha \Gamma}{\alpha^2 + m^2 \Gamma^2 \omega_0^2} 
  + \frac{t_c^2 \xi_0^2 \Omega}{\alpha^2 + m^2 \Gamma^2 \omega_0^2} \Bigg[
    \frac{
      m (\alpha + t_c^2 \alpha \Omega^2)
      + m t_c^2 \Gamma^2 (\alpha - m \omega_0^2)
      + t_c (t_c \alpha + 2 m \Gamma)(\alpha - m \omega_0^2)
    }{
      m^2 t_c^4 \Omega^4 
      + \left(m - t_c^2 \alpha + m t_c (\Gamma + t_c \omega_0^2)\right)^2
      + m t_c^2 \Omega^2 \left(2 t_c^2 \alpha 
        + m \left(2 + t_c (\Gamma (2 + t_c \Gamma) - 2 t_c \omega_0^2)\right)\right)
    } \\
\qquad\qquad\qquad\ - 
    \frac{
      m (\alpha + t_c^2 \alpha \Omega^2)
      + m t_c^2 \Gamma^2 (\alpha + m \omega_0^2)
      - t_c (t_c \alpha - 2 m \Gamma)(\alpha + m \omega_0^2)
    }{
      m^2 t_c^4 \Omega^4 
      + \left(m + t_c^2 \alpha + m t_c (\Gamma + t_c \omega_0^2)\right)^2
      + m t_c^2 \Omega^2 \left(-2 t_c^2 \alpha 
        + m \left(2 + t_c (\Gamma (2 + t_c \Gamma) - 2 t_c \omega_0^2)\right)\right)
    }
\Bigg].
\end{split}
\label{eq:omega_viscous}
\end{equation}
}
}
\end{widetext}

It is to be noted that the first term in Eq.~\eqref{eq:omega_viscous} is non-zero finite only when both $\alpha$ and $\Delta T$ are non-zero, i.e., in the presence of potential anisotropy and temperature gradient in the system, which induces particle gyration\cite{mancois2021two_temp}.
Similarly, the second term of Eq.~\eqref{eq:omega_viscous} has a finite value only for non-zero values of both $\Omega$ and $t_{c}$ and represents the contribution from the chiral self-propulsion of the particle.
Hence, the resultant rotational motion is solely due to the combined effects of both gyration and chiral self-propulsion of the particle. That is why $\langle \omega \rangle$ depends on $\alpha$, $\Delta T$, $\Omega$, and $t_{c}$. By the complex interplay of these parameters, $\langle \omega \rangle$ is expected to possess different qualitative behaviors, which we explore in the following sections.

\begin{center}
    \textit{Case 1 : Pure chiral self-propulsion ($\Delta T = 0$ or $\alpha=0$)} \\ 
\end{center}
The case $\alpha = 0$ or $\Delta T = 0$ corresponds to the pure chiral motion of the particle in a viscous medium. For $\alpha = 0$, the expression of $\langle \omega \rangle$ [Eq.~\eqref{eq:omega_viscous}] reduces to
\begin{equation}
\begin{split}
\langle \omega \rangle = 
-\frac{2 t_c^3 (2 + t_c \Gamma) \, \xi_0^2 \Omega}{
  m^2 \Gamma [
    (1 + t_c^2 \Omega^2) \left((1 + t_c \Gamma)^2 + t_c^2 \Omega^2 \right)
  ]
}
\\
\qquad \times
\frac{1}{
  [
    2 t_c^2 (1 + t_c (\Gamma - t_c \Omega^2)) \omega_0^2 
    + t_c^4 \omega_0^4
  ]
}.
\label{eq:omega_viscous_a0}
\end{split}
\end{equation}
\begin{figure*}
    \centering
        \includegraphics[width=\linewidth]{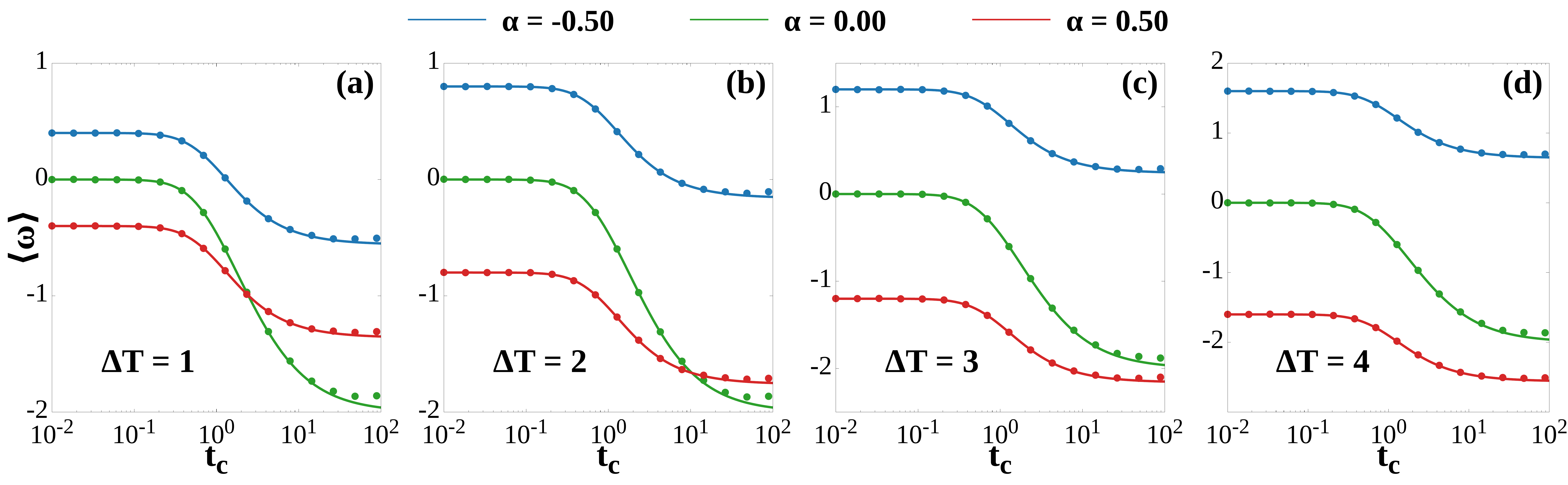}
    \caption{$\langle \omega \rangle$ [Eq.~\eqref{eq:omega_viscous}] as a function of $t_c$ for different values of $\alpha$ is shown in (a) for $\Delta T = 1$, in (b) for $\Delta T = 2$, in (c) for $\Delta T = 3$ and in (d) for $\Delta T = 4$. The other common parameters are $\Omega = \omega_0 = \Gamma = m = \xi_0 = 1$. The solid line represents the analytical results, and the dots represent the results obtained from numerical simulation.
 }
    \label{fig:omega_vs_tc_vary_a_mag}
\end{figure*}
From this equation, it is to be noted that $\langle \omega \rangle$ is non-zero finite for non-zero values of both $t_c$ and $\Omega$, reflecting that for a symmetric potential, the presence of chiral self-propulsion is responsible for having a non-zero $\langle \omega \rangle$. 
From Eq.~\eqref{eq:omega_viscous_a0}, it is also evident that $\langle \omega \rangle$ is always negative if $\Omega$ is positive. In Fig.~\ref{fig:omega_viscous_a0}(a), we have plotted $\langle \omega \rangle$ as a function of $t_c$ for different values of $\Omega$. It is observed that for a fixed $\Omega$, $\langle \omega \rangle$ shows a nonmonotonic dependence on $t_{c}$. In $t_{c} \rightarrow 0$ limit,  $\langle \omega \rangle = 0$ and for lower $t_c$ regimes, $\langle \omega \rangle \approx -\frac{4 t_c^3 \, \xi_0^2 \Omega}{m^2 \Gamma} \label{eq:omega_simple}$, which is proportional to $-t_c^3$. Hence, $\langle \omega \rangle$ decreases like $t_c^3$ in the lower $t_{c}$ regimes, and attains a minimum value.
It increases with further increase in value of $t_c$ and attains a constant value in $t_c \rightarrow \infty$ limit as
\scalebox{0.99}{ 
\parbox{\linewidth}{
\begin{equation}
{
\lim_{t_c \to \infty} \langle \omega \rangle 
= -\frac{2 \Gamma \xi_0^2 \Omega}{
m^2 \Gamma^3 \Omega^2 + m^2 \Gamma \Omega^4 
- 2 m^2 \Gamma \Omega^2 \omega_0^2 + m^2 \Gamma \omega_0^4
}
}.
\label{eq:omega_viscous_limit}
\end{equation}
}}\\
The optimum $t_{c}$ at which $\langle \omega \rangle$ shows a minimum depends on the value of $\Omega$. It increases towards zero and shifts towards the left with an increase in the value of $\Omega$.
Similarly, in Fig.~\ref{fig:omega_viscous_a0}(b), we plot $\langle \omega \rangle$ as a function of $\Omega$ for different values of $t_{c}$.
For a fixed $t_{c}$ value, as $\Omega$ increases from zero, the $\langle \omega \rangle$ first decreases with $\Omega$ (as $\langle \omega \rangle \approx -\frac{2 t_c^3 (2 + t_c \Gamma) \, \xi_0^2 \Omega}{
  m^2 \Gamma \left(1 + t_c \Gamma + t_c^2 \omega_0^2 \right)^2
}$) and approaches a minimum value. It further increases with an increase in $\Omega$ as $\langle \omega \rangle \approx -\frac{2 (2 + t_c \Gamma) \, \xi_0^2}{
  m^2 t_c \Gamma \, \Omega^3
}$ and approaches zero in the $\Omega \rightarrow \infty$ limit. The optimum $\Omega (=\Omega_0)$ at which $\langle \omega \rangle$ attains its minimum is given by
\begin{widetext}
\begin{equation}
\Omega_0 = \sqrt{
  \frac{
    -2 - 2 t_c \Gamma - t_c^2 \Gamma^2 + 2 t_c^2 \omega_0^2
  }{6 t_c^2}
  + \frac{1}{6} \sqrt{
    \frac{
      16 + 32 t_c \Gamma + 20 t_c^2 \Gamma^2 + 4 t_c^3 \Gamma^3 
      + t_c^4 \Gamma^4 + 16 t_c^2 \omega_0^2 
      + 16 t_c^3 \Gamma \omega_0^2 - 4 t_c^4 \Gamma^2 \omega_0^2 
      + 16 t_c^4 \omega_0^4
    }{t_c^4}
  }
}.
\label{eq:omega0_wide}
\end{equation}
\end{widetext}
Thus, the $\Omega_0$ depends on values of $t_c$, $\Gamma$, and $\omega_0$. By substituting $\Omega_{0}$, in Eq.~\eqref{eq:omega_viscous_a0}, we can obtain the minimum $\langle \omega \rangle$. 
From Eq.~\eqref{eq:omega_viscous_a0}, it is also evident that $\langle \omega \rangle$ is an odd function of $\Omega$ and thus the qualitative behavior of $\langle \omega \rangle$ for negative values of $\Omega$ is the same as that of its positive counterpart but with the opposite sign.
\begin{center}
    \textit{Case 2 : Pure gyration ( $\Omega = 0$ )} \\ 
\end{center}
When $\Omega = 0$, the expression for $\langle \omega \rangle$ [Eq.~\eqref{eq:omega_gen}] reduces to
\begin{equation}
    \langle \omega \rangle = \frac{-\Delta T \alpha \Gamma}{ \alpha^2 + m^2 \Gamma^2 \omega_0^2},
    \label{eq:omega_gy}
\end{equation}
which corresponds to the case of a normal Brownian gyrator~\cite{mancois2021two_temp}. As evident from Eq.~\eqref{eq:omega_gy}, $\langle \omega \rangle$ is non-zero finite only when both $\alpha$ and $\Delta T$ are non-zero. In this case, rotational motion is generated by the interplay between anisotropic harmonic confinement and the temperature gradient.
For $\alpha > 0$, the potential is effectively tilted along the $x = -y$ diagonal, while for $\alpha < 0$, the tilt is along $x = y$. This directional tilt controls the particle motion differently depending on the temperature gradient. Thus, for $\alpha < 0$ and $\Delta T > 0$ (i.e., $T_x > T_y$), the particle gyrates in a net anti-clockwise direction, resulting in a positive $\langle \omega \rangle$. Similarly, for $\Delta T < 0$, the particle gyration is clockwise, resulting in a negative $\langle \omega \rangle$.
Similarly, for $\alpha > 0$, a positive $\Delta T$ leads to negative $\langle \omega \rangle$, and a negative $\Delta T$ yields positive $\langle \omega \rangle$. Therefore, the sign of $\langle \omega \rangle$ is governed by the product $\alpha \Delta T$, being negative when $\alpha \Delta T > 0$ and positive when $\alpha \Delta T < 0$.
This highlights how steady-state gyration can emerge solely due to the complex interplay of intrinsic system anisotropies, without requiring any external torque.
\begin{center}
    \textit{Case 3: Gyration with chiral self-propulsion} \\ \centering \textit{($\Omega \neq 0$, $\alpha \neq 0$ , $t_c \neq 0$ and $\Delta T \neq 0$ )}
\end{center}

The expression for $\langle \omega \rangle$ in Eq.~\eqref{eq:omega_viscous} accounts for contributions from both gyration and chiral motion. The first term in Eq.~\eqref{eq:omega_viscous} becomes zero when either $\alpha = 0$ or $\Delta T = 0$, while the second term vanishes either for $\Omega = 0$ or $t_c = 0$.
\begin{figure}
    \centering
    \includegraphics[width=\linewidth]{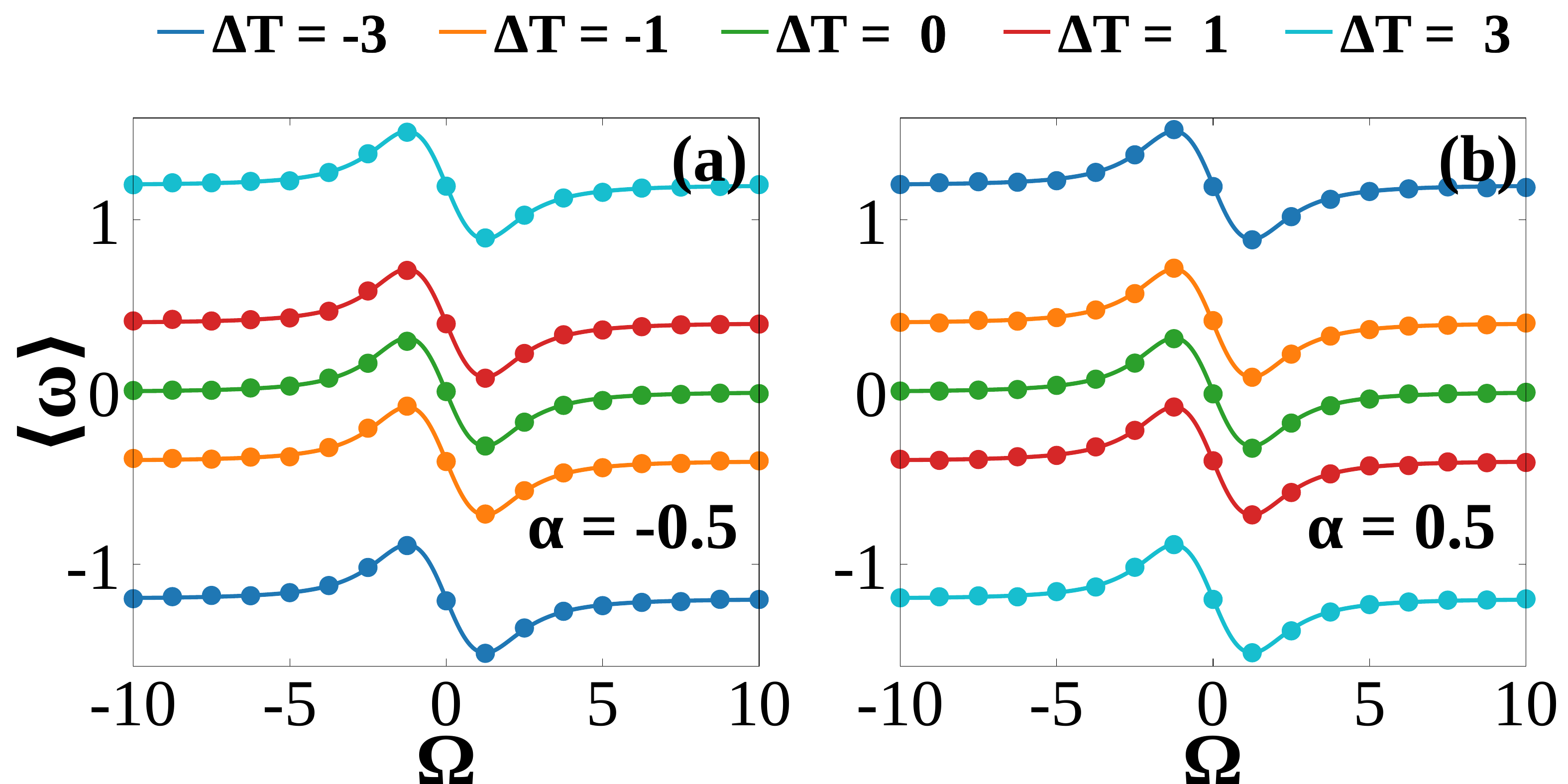}
    \caption{$\langle \omega \rangle$ [Eq.~\eqref{eq:omega_viscous}] as a function of $\Omega$ for different values of $\Delta T$ in (a) for $\alpha = -0.5$ and in (b) for $\alpha = 0.5$. Other fixed parameters are $t_c = \omega_0 = \Gamma = m = \xi_0 = 1$. The solid line represents the analytical results, and the dots represent the results obtained from numerical simulation.}
    \label{fig:omega_vs_wc_vary_Tr_mag}
\end{figure}
\begin{figure*}
    \centering
    \includegraphics[width=\linewidth]{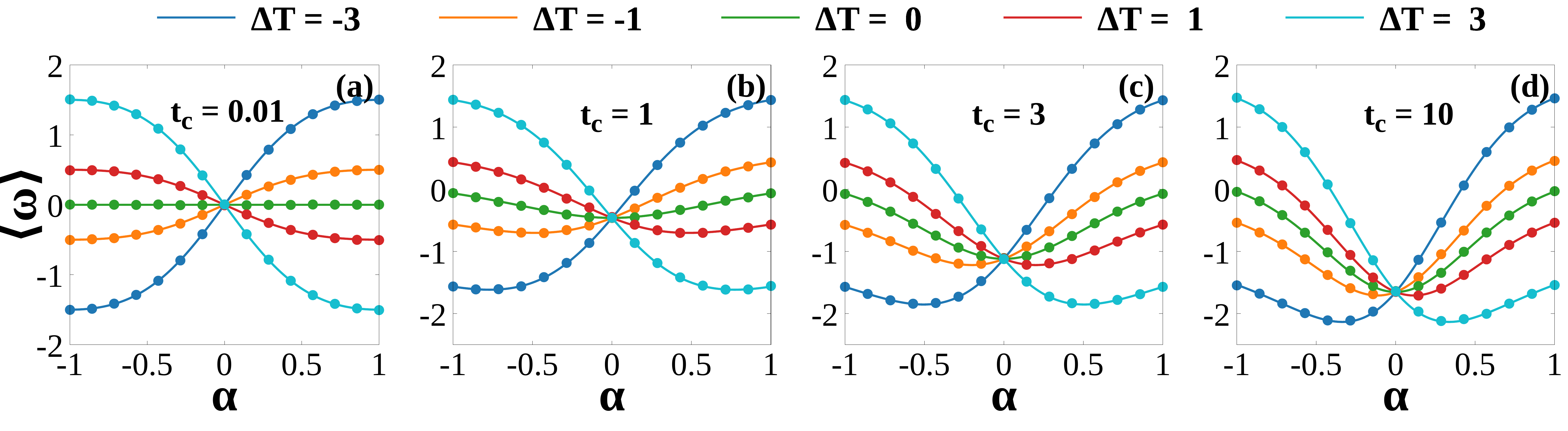}
    \caption{$\langle \omega \rangle$ [Eq.~\eqref{eq:omega_viscous}] as a function of $\alpha$ for different values of $\Delta T$ for $t_c = 0.01$ in (a), for $t_c = 1$ in (b), for $t_c = 3$, in (c), and for $t_c = 10$ in (d). The other parameters are $\Omega = \Gamma = m = \omega_0 = \xi_0 = 1$. The solid line represents the analytical results, and the dots represent the results obtained from numerical simulation.
    }
    \label{fig:omega_vs_alpha}
\end{figure*}
Figure~\ref{fig:omega_vs_tc_vary_a_mag} shows the plot of $\langle \omega \rangle$ as a function of $t_c$ for different values of $\alpha$ and for four different values of $\Delta T$ [Figs.~\ref{fig:omega_vs_tc_vary_a_mag}(a)-(d)].
It is observed that for a fixed $\alpha$, $\langle \omega \rangle$ exhibits a monotonic dependence on $t_c$ values. In the limit $t_c \to 0$, $\langle \omega \rangle$ is constant, which remains unchanged for lower  $t_c$ regimes, resulting in a plateau-like behavior. In this limit, $\langle \omega \rangle$ reduces to
Eq.~\eqref{eq:omega_gy},
which corresponds to the angular momentum of a Brownian gyrator~\cite{mancois2021two_temp}. 
This expression confirms that in the absence of activity, both the potential anisotropy and temperature gradient are essential to induce gyration. The absence of either one of them leads to zero $\langle \omega \rangle$. 
With further increase in $t_{c}$ value, $\langle \omega \rangle$ [Eq.~\eqref{eq:omega_viscous}] decreases, approaches a minimum value, and then eventually saturates to a constant value in $t_c \to \infty$ limit.
When the product $\alpha \Delta T$ is negative, the system exhibits a positive $\langle \omega \rangle$ at small $t_c$, and intriguingly, it exhibits a transition from positive to negative $\langle \omega \rangle$ with an increase in the value of $t_c$ [Fig.~\ref{fig:omega_vs_tc_vary_a_mag}(a)]. This transition occurs only for lower $\Delta T$ values [Figs.~\ref{fig:omega_vs_tc_vary_a_mag}(a) and (b)].
As $\Delta T$ is increased, the transition disappears and $\langle\omega\rangle$ remains positive throughout the entire $t_c$ regimes and shifts more into the positive region [Figs.~\ref{fig:omega_vs_tc_vary_a_mag}(c) and (d)].
Similarly, if $\alpha \Delta T$ is positive, the system shows a negative $\langle \omega \rangle$ [Fig.~\ref{fig:omega_vs_tc_vary_a_mag}(a)], but it shows no transition behavior with a change in $t_c$ values.
With an increase in $\Delta T$ values, $\langle \omega \rangle$ remains negative throughout the entire $t_c$ regimes and shifts more into the negative region [Figs.~\ref{fig:omega_vs_tc_vary_a_mag}(b)-(d)].
However, for $\alpha = 0$ there is no $\Delta T$ dependence, and $\langle\omega\rangle$ remains the same for all $\Delta T$ values.

In Fig.~\ref{fig:omega_vs_wc_vary_Tr_mag}, we show the plot of $\langle \omega \rangle$ as a function of $\Omega$ for different values of $\Delta T$ and for two different $\alpha$ values in Figs.~\ref{fig:omega_vs_wc_vary_Tr_mag} (a) and (b), respectively. For a fixed $\Delta T$, the behavior of $\langle \omega \rangle$ is non-monotonic with $\Omega$. 
$\langle \omega \rangle$ shows a minima for small positive values of $\Omega$ and a maxima for small negative values of $\Omega$.
However, $\langle \omega \rangle$ is constant and becomes independent of $\Omega$ for very large positive and negative values of $\Omega$. 
For $\Delta T = 0$, $\langle \omega \rangle$ changes sign as a function of $\Omega$. 
This is because, for $\Delta T=0$, the first term of Eq.~\eqref{eq:omega_viscous} vanishes and $\langle \omega \rangle$ becomes an odd function of $\Omega$. $\langle\omega\rangle$ is positive when $\Omega$ is negative and is negative when $\Omega$ is positive. 
From Figs.~\ref{fig:omega_vs_wc_vary_Tr_mag} (a) and (b), it is also observed that the sign of $\langle \omega \rangle$ changes with a change in the sign of the product $\alpha \Delta T$.
For large positive values of $\Delta T$, $\langle\omega\rangle$ is positive for a negative $\alpha$ value [Fig.~\ref{fig:omega_vs_wc_vary_Tr_mag} (a)]. Similarly, for large negative values of $\Delta T$, $\langle\omega\rangle$ is positive for a positive $\alpha$ value [Fig.~\ref{fig:omega_vs_wc_vary_Tr_mag} (b)].  

In Fig.~\ref{fig:omega_vs_alpha}, we plot $\langle \omega \rangle$ as a function of $\alpha$ for different values of $\Delta T$ and for four different $t_{c}$ values in Figs.~\ref{fig:omega_vs_alpha}(a)-(d), respectively. For a very small $t_c$ value [Fig.~\ref{fig:omega_vs_alpha}(a)], i.e., for the case of a Brownian gyrator~\cite{mancois2021two_temp}, there is a transition from a positive to negative $\langle\omega\rangle$ (for $\Delta T > 0$) and negative to positive $\langle\omega\rangle$ (for $\Delta T < 0$) through the $\alpha=0$ line.
This feature is evident from Eq.~\eqref{eq:omega_viscous}. For very small $t_c$ values (i.e., $t_{c} \rightarrow 0$ limit), the contribution of the second term of Eq.~\eqref{eq:omega_viscous} to the net $\langle \omega \rangle$ vanishes. Since the first term is proportional to $-\alpha \Delta T$, $\langle \omega \rangle$ is positive only when $\Delta T$ and $\alpha$ have opposite signs, and $\langle \omega \rangle$ is negative only when both $\Delta T$ and $\alpha$ have the same sign. This results in the direction reversal of rotational motion as discussed above in the $\langle \omega \rangle$ vs $\alpha$ plot [Fig.~\ref{fig:omega_vs_alpha}(a)].
With an increase in $t_{c}$ value, $\langle \omega \rangle$ shifts towards the negative region [Figs.~\ref{fig:omega_vs_alpha}(b)-(d)] due to the contribution from the second term of Eq.~\eqref{eq:omega_viscous}. 
For $t_c \gg 0$, $\langle \omega \rangle$ is negative for most of the parameter regimes. 

\begin{figure*}
    \centering
    \includegraphics[width=0.85\linewidth]{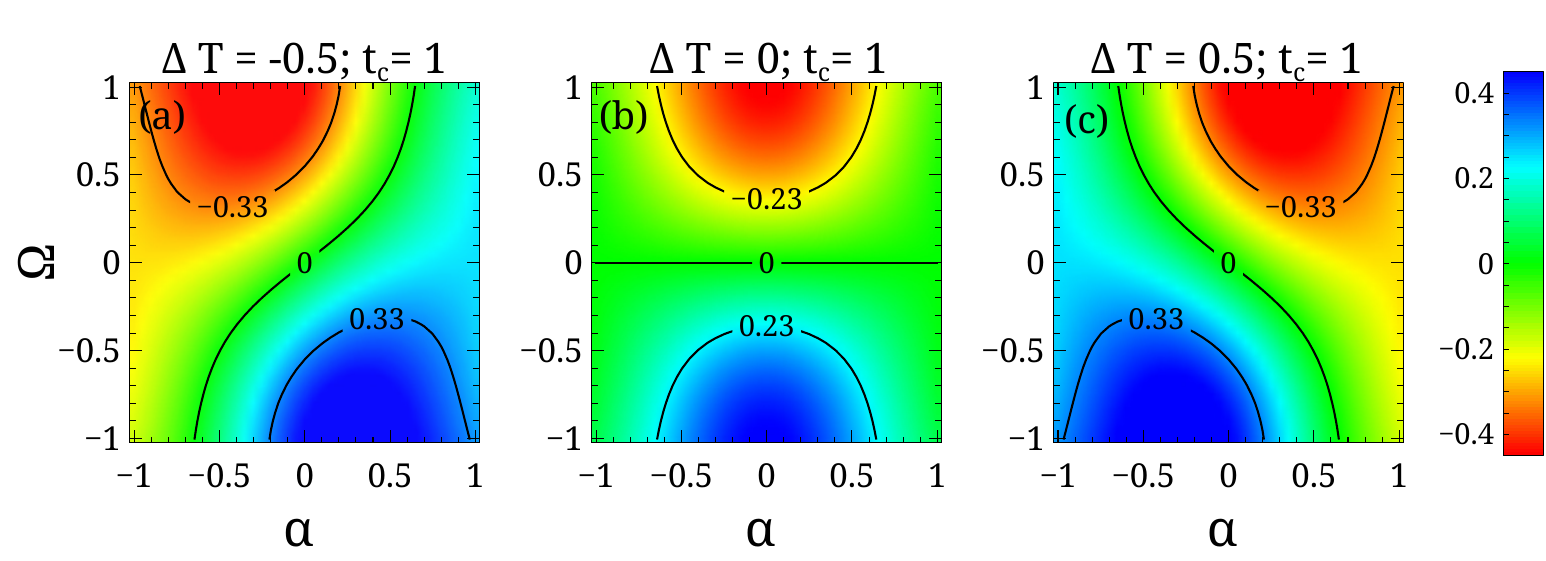}
    \caption{2d parametric plot of $\langle \omega \rangle$ [Eq.~\eqref{eq:omega_viscous}] as a function of $\Omega$ and $\alpha$ is shown in (a) for $\Delta T = -0.5$, in (b) for $\Delta T = 0$ and in (c) for $\Delta T = 0.5$. The other common parameters are $t_c = m = \Gamma = \omega_0 = \xi_0 = 1$.}
    \label{fig:omega_vs_wc_a}
\end{figure*}

In Fig.~\ref{fig:omega_vs_wc_a}, we show the 2d plot of $\langle \omega \rangle$ as a function of both $\alpha$ and $\Omega$, for three different $\Delta T$ values and for a fixed $t_c$, in Figs.~\ref{fig:omega_vs_wc_a}(a)-(c), respectively. 
The solid dark lines of each figure correspond to the contour lines of $\langle \omega \rangle$. 
It is observed that the neutral line along which $\langle \omega \rangle =0$ always passes through the origin ($\Omega = 0, \alpha=0$) of the $\Omega-\alpha$ plane. This is because this point corresponds to the case with no gyration and no chiral self-propulsion. The locus of all points on this neutral line satisfies the equation
\begin{widetext}
\begin{equation}
\begin{split}
&\frac{
  \Gamma^2 m t_c^2 (\alpha + m \omega_0^2)
  + m \alpha (1 + t_c^2 \Omega^2)
  - t_c (\alpha + m \Omega_0^2)(\alpha t_c - 2 \Gamma m)
}{
  m^2 t_c^4 \Omega^4 
  + m t_c^2 \Omega^2 \left[ m \left(t_c (\Gamma (\Gamma t_c + 2) - 2 t_c \omega_0^2) + 2\right) - 2 \alpha t_c^2 \right] 
  + \left( m t_c (\Gamma + t_c \omega_0^2) + m + \alpha t_c^2 \right)^2
} \\
&\quad - 
\frac{
  \Gamma^2 m t_c^2 (\alpha - m \omega_0^2)
  + m \alpha (1 + t_c^2 \Omega^2)
  + t_c (\alpha - m \omega_0^2)(2 \Gamma m + \alpha t_c)
}{
  m^2 t_c^4 \Omega^4 
  + m t_c^2 \Omega^2 \left[ m \left(t_c (\Gamma (\Gamma t_c + 2) - 2 t_c \omega_0^2) + 2\right) + 2 \alpha t_c^2 \right] 
  + \left( m t_c (\Gamma + t_c \omega_0^2) + m - \alpha t_c^2 \right)^2
}
= -\frac{\alpha \Gamma \Delta T}{\xi_0^2 t_c^2 \Omega}.
\end{split}
\label{eq:neutral-locus}
\end{equation}
\end{widetext}

The existence of this neutral line implies that for the parameter values on this line, both the gyration and chiral effects get canceled, resulting in a net zero $\langle \omega \rangle$. 
When $\Delta T = 0$ and for a finite $t_c$ value [Fig.~\ref{fig:omega_vs_wc_a}(b)], the equation for the neutral line [Eq.~\eqref{eq:neutral-locus}] reduces to $\Omega = 0$. In this case, $\langle \omega \rangle$ is negative for $\Omega > 0$ and vice versa. This is due to the fact that $\langle \omega \rangle$ is an odd function of $\Omega$ [Eq.~\eqref{eq:omega_viscous_a0}].
On the other hand, for $\Delta T < 0$ [Fig.~\ref{fig:omega_vs_wc_a}(a)], the neutral line crosses the $\Omega-\alpha$ parametric plane from the bottom-left to the top-right diagonal. As a result, $\langle \omega \rangle$ changes from a positive value to a negative value along the top-left to bottom-right diagonal of the $\Omega-\alpha$ parametric plane.
Similarly, for $\Delta T > 0$ [Fig.~\ref{fig:omega_vs_wc_a}(c)], the neutral line passes from the bottom-right to the top-left diagonal of the $\Omega-\alpha$ parametric plane. In this case, a shift of $\langle \omega \rangle$ from positive to negative values is observed from the bottom-left corner of the parametric plane to the top-right corner. 
It is to be noted that for both of these above two cases, the system exhibits a transition from positive to negative $\langle \omega \rangle$ (or vice-versa) in the $\alpha-\Omega$ parameter plane depending on the sign of the product of $\alpha \Delta T$. This transition is through a neutral line along which both $\Omega$ (or $t_{c}$) and $\alpha$ (or $\Delta T)$ are non-zero [Figs.~\ref{fig:omega_vs_wc_a}(a) and (c)]. Thus, the system exhibits a net zero $\langle \omega \rangle$ even for parameters that violate the equilibrium condition (i.e., $\Delta T \neq 0$ and $t_c \neq 0$). This is due to the cancellation of the contributions from both effects of gyration and chiral motion, resulting in an effective zero $\langle \omega \rangle$ along the neutral line. 

\begin{figure}
    \centering
    \includegraphics[width=0.8\linewidth]{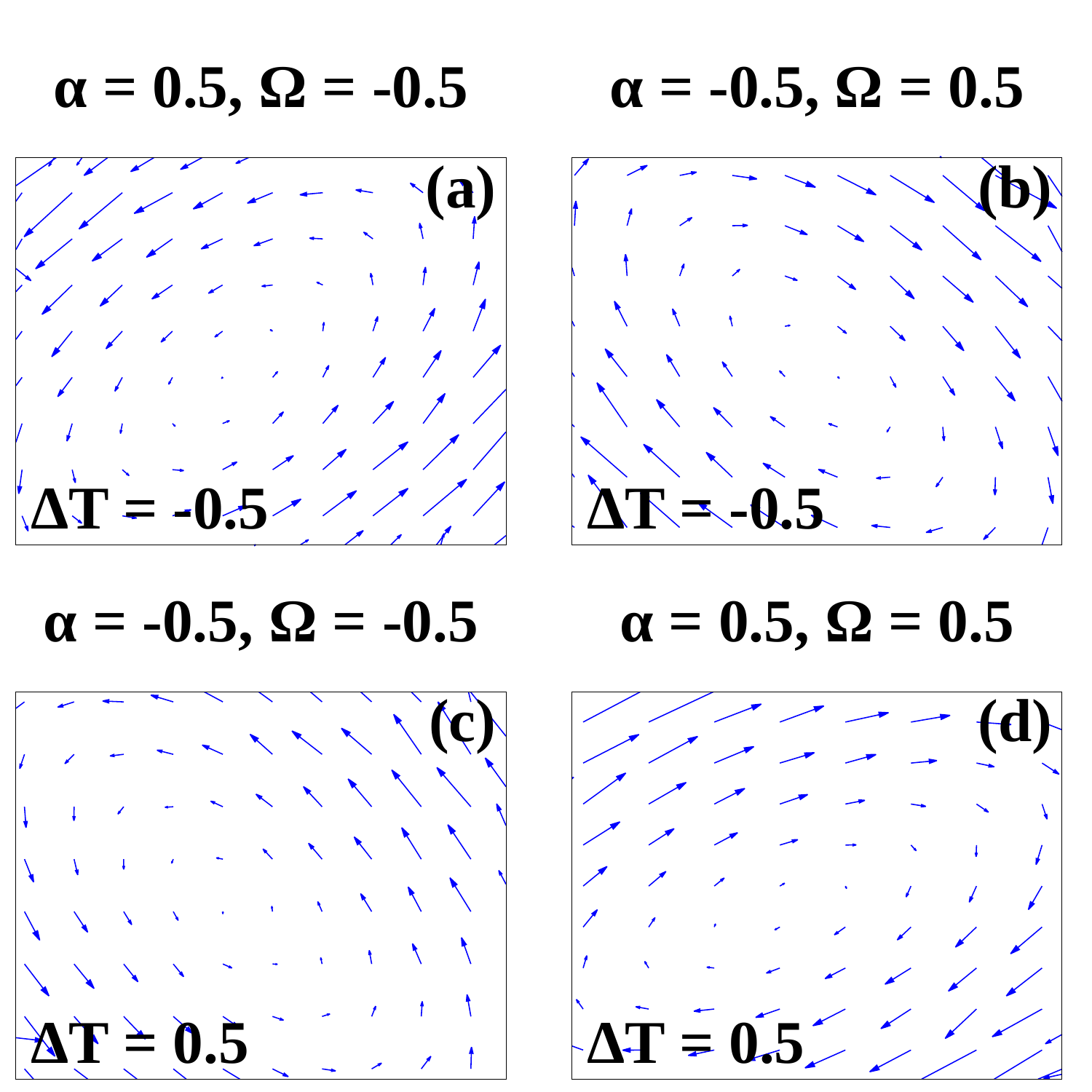}
    \caption{2d vector field plot of particle velocity flux is shown in (a) and (b) for $\Delta T = -0.5$, in (c) and (d) for $\Delta T = 0.5$.
    The other common parameters are $t_c = m = \Gamma = \omega_0 = \xi_0 = 1$. 
    }
    \label{fig:omega_vector_field}
\end{figure}

Furthermore, in order to understand the phase transition with opposite signs of $\langle \omega \rangle$ as observed in Figs.~\ref{fig:omega_vs_wc_a}(a) and (c), we simulate the velocity flux of the particle in the corresponding parameter regimes. 
The simulations were performed using the Euler-Maruyama integration scheme with a time step of $10^{-2}$ and a total of $10^{3}$ timesteps, and steady-state data were averaged over $10^{5}$ independent realizations.
In Fig.~\ref{fig:omega_vector_field}, we present the simulated vector field plots of particle velocity flux in the parameter regimes of opposite $\langle \omega \rangle$ values. It is observed that for the parameter regimes of bottom-right corner of Fig.~\ref{fig:omega_vs_wc_a}(a) with +ve $\langle \omega \rangle$, the velocity flux is in the anticlockwise direction [Fig.~\ref{fig:omega_vector_field}(a)], whereas in the parameter regimes of the top-left corner of Fig.~\ref{fig:omega_vs_wc_a}(a) with -ve $\langle \omega \rangle$, the velocity flux is in the clockwise direction [Fig.~\ref{fig:omega_vector_field}(b)]. Thus, the direction reversal from anticlockwise to clockwise directions of velocity flux of the particle [Figs.~\ref{fig:omega_vector_field}(a) and (b)] is complementing the transition from +ve to -ve $\langle \omega \rangle$ as observed in Fig.~\ref{fig:omega_vs_wc_a}(a). Similarly, for the parameter regimes of bottom-left corner of Fig.~\ref{fig:omega_vs_wc_a}(c), the velocity flux is observed to be in the anticlockwise direction [Fig.~\ref{fig:omega_vector_field}(c)] and for the parameter regime of top-right corner of Fig.~\ref{fig:omega_vs_wc_a}(c), the velocity flux is observed to be in the clockwise direction [Fig.~\ref{fig:omega_vector_field}(d)]. Hence, the directional reversal of velocity flux from anticlockwise direction [Fig.~\ref{fig:omega_vector_field}(c)] to clockwise direction [Fig.~\ref{fig:omega_vector_field}(d)] is complementing the transition from +ve to -ve $\langle \omega \rangle$ as observed in Fig.~\ref{fig:omega_vs_wc_a}(c).

\section{CONCLUSIONS}\label{sec:summary}
In this work, using both analytical approach and numerical simulation, we have explored the rotational dynamics of a CAP confined in a two-dimensional anisotropic harmonic potential and coupled to two orthogonal heat baths kept at two different temperatures. The particle is subjected to a viscoelastic environment. The system is driven out of equilibrium via two primary mechanisms: gyration induced by the combined effect of a finite temperature gradient 
and the potential anisotropy, and a chiral self-propulsion. The viscoelastic environment is realized by introducing a friction kernel to the dynamics through a generalized Langevin framework with a finite relaxation time. This allows the medium to retain the history of the particle motion. 
With these settings, we have examined the rotational dynamics of the particle by exactly deriving the average angular momentum of the particle at steady state. The angular momentum is found to be sensitive to chirality strength, activity timescale, temperature gradient, potential anisotropy, and viscoelastic memory time scale. We explore different circumstances of the rotational motion, including pure chiral self-propulsion, pure gyration, and the combined effect of both gyration and chiral self-propulsion.
Remarkably, it is observed that in the absence of a temperature gradient, the particle exhibits a memory-induced direction reversal of rotational motion.
This phenomenon is purely a consequence of the interplay between chiral activity and the memory of the medium and is absent in a purely viscous environment.
However, for such direction reversal of rotational motion to occur, potential anisotropy is also an essential ingredient, even when the temperature gradient is absent.  By introducing a temperature gradient to the system, the direction and magnitude of the angular momentum can be further modulated. Depending on the sign and strength of the temperature gradient and potential anisotropy, the system exhibits a completely clockwise, a completely anti-clockwise, or a coexistence of both phases of rotation throughout the entire activity-memory parameter space.

Moreover, in the viscous limit, direction reversal occurs only through the neutral line of chirality-potential anisotropy parameter space, along which the contributions from both gyration and chiral motion to the net angular momentum get exactly canceled.
Finally, we believe our results demonstrate that the combined effects of viscoelastic memory, chiral activity, and system anisotropies offer a robust mechanism for directional control in active matter systems. 
These findings offer a theoretical framework that could guide future biomedical experiments on micro-swimmers in bio-fluids, where memory effects and chiral dynamics play a central role~\cite{Mao2014}. 
It would also be further interesting to extend this work by including hydrodynamic interactions~\cite{Navarro2015hydrodynamics} and  external magnetic fields~\cite{muhsin2024magneto}, which might help the engineering of functional active rotors or directional switches.

\section{Acknowledgement}

We acknowledge the departmental computational facility at the Department of Physics, University of Kerala, SERB-SURE grant (SUR/2022/000377), and CRG grant (CRG/2023/002026) from DST, Govt. of India for financial support.

\begin{widetext}
\appendix
\section{} \label{sec:app_A}

For the calculation of angular momentum, we first express the dynamics Eq.~\eqref{eq:model} in matrix form. The thermal noise $\boldsymbol{\eta}(t)$ in Eq.~\eqref{eq:model} can be taken as the sum of two independent Gaussian noises, $\boldsymbol{\eta}(t)=\boldsymbol{\eta_1}(t) + \boldsymbol{\eta_2}(t)$. Here $\boldsymbol{\eta_1}(t)$ is a white noise with the correlation $\langle{\eta_1}_i(t){\eta_1}_j(t)\rangle~=~\delta_{ij}\gamma k_BT_i \delta(t - t')$, and $\boldsymbol{\eta_2}(t)$ is a colored noise with the correlation $\langle{\eta_2}_i(t){\eta_2}_j(t)\rangle~=~\delta_{ij}\frac{\gamma}{2t_c'}k_BT_i e^{-(t - t')/t_c'}$. The noise $\boldsymbol{\eta_2}(t)$ can be modeled as an Ornstein-Uhlenbeck process as
\begin{equation}
    t_c'\boldsymbol{\dot{\eta_2}}(t) = -\boldsymbol{\eta_2}(t) + \begin{pmatrix}
    \sqrt{\gamma k_B T_x} & 0 \\
    0 & \sqrt{\gamma k_B T_y}
    \end{pmatrix} \cdot \boldsymbol{\zeta}'(t),
\end{equation}
with $\boldsymbol{\zeta}'(t)$ a delta-correlated white noise. To rewrite Eq.~\eqref{eq:model} as a matrix equation, we introduce the vector ${\bf U}$ such that
\begin{equation}
    {\bf U} = \int\limits_{-\infty}^{t} \frac{\Gamma}{2t_c'} e^{-\frac{t - t'}{t_c'}} {\bf \dot{r}}(t')\; dt' 
\end{equation}
Hence, the Eq.~\eqref{eq:model} can be expressed as
\begin{equation}
    \dot{\boldsymbol{\chi}} = A\boldsymbol{\chi} + B \boldsymbol{\eta'},
    \label{eq:dynamics-matrix}
\end{equation}
with the matrices $\boldsymbol{\chi}$, $A$, $B$, and $\boldsymbol{\eta'}$ given by
\begin{equation}
    \boldsymbol{\chi} = (x\ y\ v_x\ v_y\ U_x\ U_y\ \xi_x\ \xi_y\ \eta_{2x}\ \eta_{2y})^T,
\end{equation}

\begin{equation}
    A = \begin{pmatrix}
        0 & 0 & 1 & 0 & 0 & 0 & 0 & 0 & 0 & 0 \\
0 & 0 & 0 & 1 & 0 & 0 & 0 & 0 & 0 & 0 \\
-\omega_0^2 & -\tfrac{\alpha}{m} & -\tfrac{\Gamma}{2} & 0 & - 1 & 0 & \tfrac{1}{m} & 0 & \sqrt{\tfrac{\Gamma k_B T_x}{m}} & 0 \\
-\tfrac{\alpha}{m} & -\omega_0^2 & 0 & -\tfrac{\Gamma}{2} & 0 & - 1 & 0 & \tfrac{1}{m} & 0 & \sqrt{\tfrac{\Gamma k_B T_y}{m}} \\
0 & 0 & \tfrac{\Gamma}{2 t_c'} & 0 & -\tfrac{1}{t_c'} & 0 & 0 & 0 & 0 & 0 \\
0 & 0 & 0 & \tfrac{\Gamma}{2 t_c'} & 0 & -\tfrac{1}{t_c'} & 0 & 0 & 0 & 0 \\
0 & 0 & 0 & 0 & 0 & 0 & -\tfrac{1}{t_c} & \Omega & 0 & 0 \\
0 & 0 & 0 & 0 & 0 & 0 & -\Omega & -\tfrac{1}{t_c} & 0 & 0 \\
0 & 0 & 0 & 0 & 0 & 0 & 0 & 0 & -\tfrac{1}{t_c'} & 0 \\
0 & 0 & 0 & 0 & 0 & 0 & 0 & 0 & 0 & -\tfrac{1}{t_c'}
    \end{pmatrix},
\end{equation}

\begin{equation}
    B = \begin{pmatrix}
        0 & 0 & 0 & 0 & 0 & 0 & 0 & 0 & 0 & 0 \\
0 & 0 & 0 & 0 & 0 & 0 & 0 & 0 & 0 & 0 \\
0 & 0 & \tfrac{\Gamma k_B T_x}{m} & 0 & 0 & 0 & 0 & 0 & 0 & 0 \\
0 & 0 & 0 & \tfrac{\Gamma k_B T_y}{m} & 0 & 0 & 0 & 0 & 0 & 0 \\
0 & 0 & 0 & 0 & 0 & 0 & 0 & 0 & 0 & 0 \\
0 & 0 & 0 & 0 & 0 & 0 & 0 & 0 & 0 & 0 \\
0 & 0 & 0 & 0 & 0 & 0 & \tfrac{2 \xi_0^2}{t_c} & 0 & 0 & 0 \\
0 & 0 & 0 & 0 & 0 & 0 & 0 & \tfrac{2 \xi_0^2}{t_c} & 0 & 0 \\
0 & 0 & 0 & 0 & 0 & 0 & 0 & 0 & \tfrac{1}{t_c'^2} & 0 \\
0 & 0 & 0 & 0 & 0 & 0 & 0 & 0 & 0 & \tfrac{1}{t_c'^2}
    \end{pmatrix},
\end{equation}
and 
\begin{equation}
    \boldsymbol{\eta'} = (0\ 0\ \eta_{1x}\ \eta_{1y}\ 0\ 0\ \zeta_{x}\ \zeta_{y}\ \zeta_{x}'\ \zeta_{y}')^T.
\end{equation}
Introducing the correlation matrix $\boldsymbol{\Xi}$ with elements 
\begin{equation}
    [\Xi_{i,j}] = \langle \chi_i \chi_j \rangle - \langle \chi_i \rangle \langle \chi_j \rangle.
\end{equation}
As per the correlation matrix formalism, it can be shown that $\boldsymbol{\Xi}$ satisfies the following equation:
\begin{equation}
    A\cdot \boldsymbol{\Xi} + \boldsymbol{\Xi} \cdot A^T + B = 0.
    \label{eq:Xi_relation}
\end{equation}
Once the matrix relation Eq.~\eqref{eq:Xi_relation} is solved, the magnitude of the angular momentum Eq.~\eqref{eq:M_def} can be calculated as
\begin{equation}
\begin{aligned}
    \langle \omega \rangle& = 
    \lim_{t\to \infty} (\langle v_yx \rangle - \langle v_x y \rangle). \\ 
    &= (\Xi_{4,1} - \Xi_{3,2}).
    \label{eq:omega_relation}
\end{aligned}
\end{equation}
Substituting the values of $\Xi_{4,1}$ and $\Xi_{3,2}$, we obtain the average angular momentum as reported in Eq.~\eqref{eq:omega_gen}.

\section{}\label{sec:app_B}

The $S$, $S'$, $D$, $E$, $M$, and $N$ appearing in Eq.~\eqref{eq:omega_gen} is given by

\[
\begin{aligned}
S =&\ 4 E ( m + m t_c^2 \Omega^2 )^2 
+ m^2 t_c \Gamma ( 1 + t_c^2 \Omega^2 ) 
( (2 t_c + t_c') ( 4 (t_c + t_c') + t_c (2 t_c + t_c') \Gamma ) \nonumber \\
& + t_c^2 t_c' ( 4 t_c' + t_c (-4 + t_c' \Gamma) ) \Omega^2 ), \\[5pt]
S' =&\ 4 m t_c^2 ( 2 (t_c + t_c')^2 - 2 t_c^3 (t_c + 2 t_c') \Omega^2 
- 2 t_c^4 t_c'^2 \Omega^4 
+ t_c (t_c + t_c') \Gamma ( t_c' + t_c ( 2 + t_c t_c' \Omega^2 ) ) ), \\
D =&\ 8 {t'_c}^4 \alpha^4 
+ m^2 \alpha^2 ( 
(-2 + t'_c \Gamma) (2 + t'_c \Gamma)^2 
+ 2 {t'_c}^2 (-8 + t'_c \Gamma (-6 + t'_c \Gamma)) \omega_0^2 
- 8 {t'_c}^4 \omega_0^4 
) \nonumber\\
& - 2 m^4 \Gamma^2 \omega_0^2 ( 
2 + t'_c (\Gamma + t'_c \omega_0^2) 
)^2, \\
\hspace{2em}
E =&\ (t_c + t_{c}')^2 + t_c^2 t_{c}'^2 \Omega^2, \\
M =&\ 16 m t_c^2 t_{c}'^4 \alpha^4 E 
+ 4 t_{c}'^3 \alpha^3 F 
+ 2 t_c \alpha^2 G 
+ \alpha H + I, \\[5pt]
N =&\ 16 m t_c^2 t_{c}'^4 \alpha^4 E 
- 4 t_{c}'^3 \alpha^3 F 
+ 2 t_c \alpha^2 G 
- \alpha H + I. \\[5pt]
\text{Where,} \nonumber \\[3pt]
\hspace{2em}
F =&\ m ( 
2 m t_c \Gamma ( t_c^3 - 2 t_c^2 t_c' - 3 t_c t_c'^2 - 2 t_c'^3 + t_c^2 ( t_c - 2 t_c' ) t_c'^2 \Omega^2 ) 
- E m t_c' ( 4 + t_c^2 ( \Gamma^2 + 4 \Omega^2 - 4 \omega_0^2 ) ) 
), \\
\hspace{2em}
G =&\ m^2 ( 
-8 E m t_c ( 1 + t_c'^2 \omega_0^2 )^2 
+ m t_c' \Gamma ( 
-4 t_c'^3 ( 2 + t_c' \Gamma ) 
- t_c t_c'^2 ( 12 + t_c' \Gamma ( 6 + t_c' \Gamma ) ) 
- 4 t_c^2 t_c' ( 4 + t_c' ( \Gamma + t_c' ( 2 + t_c' \Gamma ) \Omega^2 ) ) 
\nonumber\\
& + t_c^3 ( -4 + t_c' ( \Gamma ( 2 + t_c' \Gamma ) - t_c' ( 12 + t_c' \Gamma ( 6 + t_c' \Gamma ) ) \Omega^2 ) ) 
- 2 t_c'^2 ( 4 t_c'^3 + t_c t_c'^2 ( 14 + t_c' \Gamma ) + 2 t_c^2 t_c' ( 8 + t_c' ( \Gamma + 2 t_c' \Omega^2 ) ) \nonumber\\
& + t_c^3 ( 2 + t_c' ( \Gamma + t_c' ( 6 + t_c' \Gamma ) \Omega^2 ) ) ) \omega_0^2
) 
), \\
\hspace{2em}
H =&\ m^4 ( 
\Gamma ( 2 t_c'^2 + 2 t_c^2 ( 2 + t_c' \Gamma ) + t_c t_c' ( 4 + t_c' \Gamma ) ) 
( 8 t_c' + ( 2 + t_c \Gamma ) ( 4 t_c + t_c' ( 2 t_c + t_c' ) \Gamma ) ) 
+ t_c^2 t_c'^2 \Gamma \Omega^2 ( 8 t_c' ( 4 + t_c' \Gamma ) \nonumber\\
& + 4 t_c ( 12 + t_c' \Gamma ( 6 + t_c' \Gamma ) ) + t_c^2 ( \Gamma ( -8 + t_c'^2 \Gamma^2 ) + 4 t_c' ( 4 + t_c' \Gamma ) \Omega^2 ) ) 
+ 2 ( 4 t_c'^4 ( 4 + 3 t_c' \Gamma )\nonumber\\
& + 4 t_c t_c'^3 ( 2 + t_c' \Gamma ) ( 4 + 3 t_c' \Gamma ) + 2 t_c^3 t_c' ( 2 + t_c' \Gamma ) ( -4 + t_c' ( 4 + 3 t_c' \Gamma ) ( \Gamma + 2 t_c' \Omega^2 ) ) \nonumber\\
& + t_c^2 t_c'^2 ( 8 + t_c' ( \Gamma ( 40 + t_c' \Gamma ( 20 + 3 t_c' \Gamma ) ) + 8 t_c' ( 4 + 3 t_c' \Gamma ) \Omega^2 ) ) + t_c^4 ( -8 + t_c' ( \Gamma ( -4 + 5 t_c' \Gamma ( 2 + t_c' \Gamma ) ) \nonumber\\
& + t_c' ( -2 + t_c' \Gamma ) ( -4 + t_c' \Gamma ( 2 + 3 t_c' \Gamma ) ) \Omega^2 + 4 t_c'^3 ( 4 + 3 t_c' \Gamma ) \Omega^4 ) ) ) ) \omega_0^2 \nonumber\\
& - 16 t_c t_c'^3 \Gamma ( t_c^3 + 2 t_c^2 t_c' - t_c'^3 + t_c^2 ( 2 t_c - t_c' ) t_c'^2 \Omega^2 ) \omega_0^4 
+ 4 E ( 4 + 4 t_c'^4 \omega_0^4 + t_c^2 ( 4 \Omega^2 + t_c'^2 ( -8 + t_c'^2 ( \Gamma^2 + 4 \Omega^2 ) ) \omega_0^4 \nonumber\\
& - 4 t_c'^4 \omega_0^6 ) )
), \\
\hspace{2em}
I =&\ 4 m^5 t_c \Gamma \omega_0^2 ( 2 + t_c' ( \Gamma + t_c' \omega_0^2 ) ) ( 
\Gamma ( 4 t_c'^3 + t_c t_c'^2 ( 6 + t_c' \Gamma ) + t_c^3 ( 2 + t_c' \Gamma ) ( 1 + t_c'^2 \Omega^2 ) + t_c^2 t_c' ( 4 + t_c' ( \Gamma + 4 t_c' \Omega^2 ) ) ) 
\nonumber\\
& + 4 t_c'^2 ( ( t_c + t_c' )^2 + t_c^2 t_c'^2 \Omega^2 ) \omega_0^2 
+ E ( 4 + t_c t_c'^2 \Gamma \omega_0^2 ) 
).
\end{aligned}
\]

\end{widetext}

\end{document}